# Dense Optical Flow Estimation Using Sparse Regularizers from Reduced Measurements

**Muhammad Wasim Nawaz[1], Abdesselam Bouzerdoum[2], Senior Member, IEEE, Muhammad Mahboob Ur Rahman[3], Member, IEEE, Ghulam Abbas[4], Senior Member, IEEE, Faizan Rashid[5]**

[1] Department of Computer Engineering, The University of Lahore 54000, Pakistan
[2] College of Science and Engineering, Hamad Bin Khalifa University (HBKU), Qatar.
[3] Department of Electrical and Computer Engineering, KAUST 23955, Saudi Arabia
[4] Department of Electrical Engineering, The University of Lahore 54000, Pakistan
[5] The University of Doha for Science and Technology, Al Tarafa, Jelaiah Street, Duhail North, Doha, Qatar

Corresponding author: Faizan Rashid (e-mail: Faizan.rashid@udst.edu.qa).

This paragraph of the first footnote will contain support information, including sponsor and financial support acknowledgment.

**ABSTRACT** Optical flow is the pattern of apparent motion of objects in a scene. The computation of optical flow is a critical component in numerous computer vision tasks such as object detection, visual object tracking, and activity recognition. Despite a lot of research, efficiently managing abrupt changes in motion remains a challenge in motion estimation. This paper proposes novel variational regularization methods to address this problem since they allow combining different mathematical concepts into a joint energy minimization framework. In this work, we incorporate concepts from signal sparsity into variational regularization for motion estimation. The proposed regularization uses robust $\ell 1$ norm, which promotes sparsity and handles motion discontinuities. By using this regularization, we promote the sparsity of the optical flow gradient. This sparsity helps recover a signal even with just a few measurements. We explore recovering optical flow from a limited set of linear measurements using this regularizer. Our findings show that leveraging the sparsity of the derivatives of optical flow reduces computational complexity and memory needs.

## I. INTRODUCTION

Optical flow computation, a fundamental task in computer vision, involves estimating motion from digital videos. Optical flow methods provide an approximation of the movement between consecutive image frames. Accurate optical flow estimates are essential for the extraction of meaningful information from video sequences. These precise estimates play a critical role in activities such as object detection and tracking, surveillance, motion compensation coding, depth measurement, and video analysis. Therefore, it is crucial to acquire precise optical flow estimations.

Estimating optical flow is a challenging problem due to the well-known aperture problem [1], which is based on the fact that when moving objects are viewed locally in a digital image sequence, the direction of motion of local features or objects is ambiguous. Variational methods help address the aperture problem by regularizing the optical flow field. These methods try to estimate optical flow by introducing extra constraints such as the smoothness constraint, i.e., the motion of the neighboring pixels in a digital video changes smoothly over time. This ensures that the estimated optical flow is more accurate and reflects how things are truly moving in the video. Hence, the regularization enhances the stability of solutions to get realistic estimates of the motion field.

Regularization of optical flow is typically accomplished using the $\ell_1$ norm since it makes the regularizers convex, and efficient convex optimization techniques can be employed to minimize it. The use of the $\ell_1$ norm produces sparse solutions by penalizing the small magnitudes. It also preserves large magnitudes that usually occur at object boundaries and motion discontinuities. The Total variation (TV) is an example of a convex regularizer that utilizes the $\ell_1$ norm with the gradient magnitude [2]. Since motion discontinuities usually occur at object boundaries, a TV regularizer handles discontinuities in optical flow better than a regularizer that uses an $\ell_2$ norm.

Optical flow estimation using TV regularization produces the sparse gradient flow field. However, a TV regularizer performs poorly in areas of low texture, producing unwanted staircase-like artifacts. Therefore, this paper suggests a new sparsity enhancing regularizer that aims to overcome the drawbacks of the TV regularizers. We apply this proposed regularizer for estimating optical flow from a small number of intensity derivative measurements. We conduct thorough experiments, demonstrating that optical flow estimation using the proposed regularizer generally yields lower mean end point error compared to local TV regularizers.

This paper's structure is as follows. Section 2 discusses the relevant literature on sparse regularization for optical flow estimation. Section 3 gives details of TV regularizers that promote sparsity of the gradient of unknown signals. Section 4 introduces a novel sparse regularizer for optical flow estimation. Section 5 integrates the proposed regularizer into an energy minimization framework using numerous optical flow data terms. It also gives a fast and convex algorithm for estimating optical flow from the combined energy of the regularization and data terms. Section 6 outlines the experimental results on synthetic and real video sequences, showcasing the effectiveness of the proposed method. Section 7 concludes the paper.

## II. LITERATURE REVIEW

Variational methods stabilize challenging problems by adding extra constraints such as a smoothness constraint on unknown signals. They provide a balance between matching the data and reducing the norms of the solution. Horn and Schunck's [1] suggested optical flow estimation using Tikhonov regularization [3], a method that uses Euclidean distance or $\ell_2$ norm, which tends to smooth out motion boundaries.

To retain the sharpness of motion boundaries, regularization needs to adapt to changes in motion. Often, motion boundaries align with object boundaries in digital videos, resulting in significant image gradients at these points. A concept called oriented smoothness was used, adapting smoothness based on motion boundary orientation [4]. Robust penalty functions were also introduced for the regularization terms to reduce blurring of boundaries [5, 6]. This shift from quadratic regularization to using $\ell_1$ norm-based penalty functions preserved sharp motion boundaries and encouraged sparsity of the gradient of optical flow [6, 7].

Total variation (TV) and its variants are applied to regularize optical flow in several works [2, 4, 5, 7, 8, 9]. TV promotes a piecewise smooth optical flow whose gradient is sparse, thereby preserving motion boundaries better compared to quadratic regularization. However, regularizing with homogeneous TV can smooth edges and strong intensity regions that aren't strictly horizontal or vertical [2]. Given that most motion boundaries align with image discontinuities, adapting the regularization to the structure of the image offers better preservation of these boundaries. To achieve this, image-adaptive TV has been used [4, 8].

Anisotropic and isotropic TV regularizers compute intensity derivatives of an image, assuming the image consists of connected objects that are smooth within. However, they struggle to retain texture and fine details of objects, particularly in low-textured regions of the video sequences. Using multiple image intensity gradients in a small neighborhood with a nonlocal TV has been effective in mitigating these issues [7, 9]. Nonetheless, nonlocal TV tends to be slower, especially for larger nonlocal window sizes.

In variational methods, the matching term often struggles with outliers when it is minimized using a quadratic penalty. To address this, the convex $\ell_1$ norm and Charbonnier penalty are employed with the data terms in variational optical flow methods [10]. Furthermore, an optical flow constraint (OFC) based data term struggles to handle scenarios involving changes in brightness. When intensity constancy assumption is violated, intensity derivatives are affected less than the intensity itself. As a result, a data term using the gradient constancy has been proposed [11].

The OFC and the gradient constancy are linearized data terms that perform poorly when there is a large inter-frame motion of the objects. Thus, large displacement optical flow has been proposed using dense correspondence [12] and descriptor matching [13].

Convolutional Neural Networks (CNNs) have gained attention for unsupervised motion estimation in optical flow [14, 15]. An unsupervised optical flow method using bidirectional census loss has also been introduced [16]. FlowNet, an end-to-end CNN framework, is proposed to estimate optical flow [17]. Deep neural networks for large displacement optical flow have also been introduced [18].

Attention based approaches for optical flow estimation have been reported in [32, 33, 34]. The method presented in [32] incorporates overlapping attention mechanisms to enhance global matching, resulting in an improved optical flow accuracy. Other work presented in [33], offers a survey on optical flow and scene flow estimation. The paper [34] employs a cross-attentional flow transformer architecture to enhance the accuracy and robustness of optical flow estimation.

In this paper, we use the traditional variational energy minimization framework for optical flow estimation. We do not employ any deep learning technique as it is beyond the scope of the presented work.

## III. TOTAL VARIATION REGULARIZATION

TV regularization was initially applied to denoise images in [19]. Since then, TV and its variations have been widely adopted in many ill-posed vision tasks. This section introduces anisotropic and isotropic TV regularizers, emphasizing their role in promoting sparsity.

## A. ISOTROPIC TV

An isotropic quantity remains constant irrespective of the direction. Similarly, an isotropic regularizer enforces consistent regularization in all directions. Consider a digital image $F$ defined over a domain $\mathfrak{D}$ with horizontal and vertical coordinates $x$ and $y$, respectively. The discrete isotropic TV of $F$ is defined as the sum of the image gradient magnitude at each pixel:

$$TV_{L2} = \sum_{i,j \in \mathfrak{D}} |\nabla F(i,j)| = \sum_{i,j \in \mathfrak{D}} \left[ \sqrt{\nabla_x F(i,j)^2 + \nabla_y F(i,j)^2} \right] \quad (1)$$

The summation of gradient magnitudes in isotropic TV gives it a semi-norm property. The $\ell_1$ norm of gradients encourages the sparsity of the recovered image in the gradient domain. Hence, minimizing isotropic TV results in a piecewise smooth image. While isotropic TV preserves horizontal and vertical edges, it causes the smoothing of the edges and strong intensity regions that deviate from 0° or 90° angles because it minimizes the gradient magnitude. To address this, variants of TV, such as anisotropic TV is proposed.

## B. ANISOTROPIC TV

Anisotropic TV restricts the smoothing across strong intensity structures. We define the anisotropic TV for a discrete image $F$ as the sum of the absolute differences of the partial image intensity derivatives.

$$TV_{L1} = \sum_{i,j \in \mathfrak{D}} \left[ |\nabla_x F(i,j)| + |\nabla_y F(i,j)| \right] \quad (2)$$

Anisotropic TV regularization outperforms isotropic TV, especially in preserving strong intensity structures like edges and object boundaries. The object boundaries cause discontinuities in an image resulting in strong image gradients. Therefore, adapting the regularization to the structure of the image can better preserve these sharp boundaries compared to a non-adaptive regularization. In this regard, anisotropic TV regularization can be used along with an image-dependent weight function $w(|\nabla I|)$ as

$$TV_{wL1} = \sum_{\{i,j \in \mathfrak{D}\}} w(|\nabla I|) \left\{ |\nabla_x F(i,j)| + |\nabla_y F(i,j)| \right\} \quad (3)$$

For small positive numbers $\alpha$ and $\beta$, the weight function $w(|\nabla I|)$ can expressed by

$$w(|\nabla I|) = \exp\left(-\alpha |\nabla I|^\beta \right) \quad (4)$$

In contrast to the isotropic TV, the anisotropic TV can be minimized easily. As a consequence of this, a number of convex minimization strategies, such as iterative shrinkage and primal-dual methods, can be utilized for the same purpose.

## IV. PROPOSED SPARSE REGULARIZER

In this section, a novel regularizer is introduced that enhances the sparsity of the optical flow field in the gradient domain. This novel regularizer extends the variational regularizer presented in [20], which enforces intensity continuity of partial image derivatives within small neighborhoods surrounding each pixel. The proposed regularizer is designed to work with optical flow vectors. It reduces the TV-generated staircase artifacts in low-textured areas.

Let $\mathbf{v} = [v_x \ v_y]^T$ be the optical flow vector at each image pixel. We use forward differences to give the discrete partial derivatives of $\mathbf{v}$ at pixel location $(i,j)$:

$$\nabla_{x(i,j)} \mathbf{v}(i,j) = \mathbf{v}(i+1,j) - \mathbf{v}(i,j) \quad (5)$$

and

$$\nabla_{y(i,j)} \mathbf{v}(i,j) = \mathbf{v}(i,j+1) - \mathbf{v}(i,j) \quad (6)$$

We consider a 2×2 neighborhood in each component of $\mathbf{v}$ for the continuity of the partial derivatives. Partial derivatives $\nabla_{x(i,j)}$ and $\nabla_{y(i,j)}$ are continuous along all directions except their own because discontinuities in these directions help maintain sharp boundaries. As an example, $\nabla_{x(i,j)}$ enforces continuity in all directions except the horizontal one. The continuity of the partial derivative is dependent upon the boundary's associated direction. In a 2×2 neighborhood of a discrete image, there are four possible boundary directions. The constraints for the continuity of various boundary directions in this small neighborhood are outlined below:

- Horizontal, $\nabla_{yy(i,j)} = \nabla_{y(i+1,j)} - \nabla_{y(i,j)} = 0$
- Vertical, $\nabla_{xx(i,j)} = \nabla_{x(i,j+1)} - \nabla_{x(i,j)} = 0$
- $45^o$ Diagonal, $\nabla_{xy(i,j)} = \nabla_{x(i,j)} - \nabla_{y(i+1,j)} = 0$
- $135^o$ Diagonal, $\nabla_{yx(i,j)} = \nabla_{y(i,j)} - \nabla_{x(i,j)} = 0$

Figure 1 depicts these 4 boundaries and the associated flow derivatives in a 2×2 neighborhood. For all four possible boundaries, we enforce the directional continuity of partial derivatives by minimizing $\nabla_x$, $\nabla_y$, $\nabla_{xx(i,j)}$, $\nabla_{yy(i,j)}$, $\nabla_{xy(i,j)}$, and $\nabla_{yx(i,j)}$. We minimize the $\ell_1$ norm of these specified partial derivatives to regularize $\mathbf{v}$.

$$E_{reg}(\mathbf{v}) = \|\nabla_x \mathbf{v}\|_1^2 + \|\nabla_y \mathbf{v}\|_1^2 + \|\nabla_{xy} \mathbf{v}\|_1^2 + \|\nabla_{yx} \mathbf{v}\|_1^2 \\ + \|\nabla_{xx} \mathbf{v}\|_1^2 + \|\nabla_{yy} \mathbf{v}\|_1^2 \quad (7)$$

A detailed examination of the continuity constraints reveals the following equation:

$$\nabla_{xx} \mathbf{v} = \mathbf{v}(i+1,j+1) + \mathbf{v}(i,j) \\ - \mathbf{v}(i+1,j) - \mathbf{v}(i,j+1) = \nabla_{yy} \mathbf{v} \quad (8)$$

Given the objective to estimate sparse partial flow derivatives, when $\nabla_{xx(i,j)} = 0$ or $\nabla_{yy(i,j)} = 0$, it implies zero horizontal or vertical partial derivatives, respectively. This is equivalent to $\nabla_{x(i,j)} = 0$ or $\nabla_{y(i,j)} = 0$. In such cases, minimizing $||\nabla_{xx}\mathbf{v}||_1$ and $||\nabla_{yy}\mathbf{v}||_1$ is ensured by minimizing $||\nabla_x\mathbf{v}||_1$ and $||\nabla_y\mathbf{v}||_1$, respectively. Thus, we omit these two terms in Equation (1), and denote this regularizer as HVD($\mathbf{v}$):

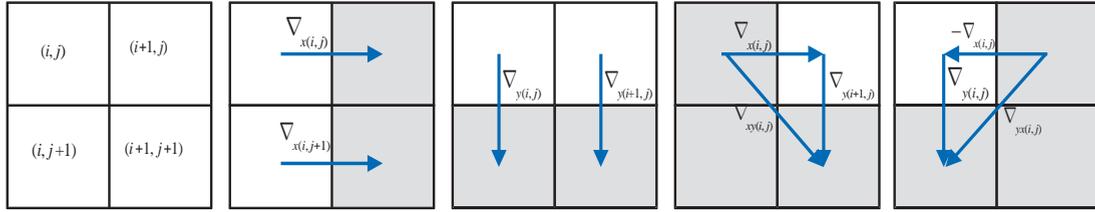

Figure 1: The continuity of derivatives in (a) 4 neighboring pixels, (b)-(e) edges in vertical, horizontal, 45° diagonal, and 135° diagonal directions, respectively. In (b)–(e), for each boundary direction, the required derivatives are indicated in blue.

$$E_{reg}(\mathbf{v}) = \text{HVD}(\mathbf{v}):$$
$$= \|\nabla_x \mathbf{v}\|_1^2 + \|\nabla_y \mathbf{v}\|_1^2 + \|\nabla_{xy}\mathbf{v}\|_1^2 + \|\nabla_{yx}\mathbf{v}\|_1^2 \quad (9)$$

$$E_{reg}(\mathbf{v}) = \text{HVD}(\mathbf{v}):$$
$$= \left\|\sqrt{(\nabla_x v_x)^2 + (\nabla_x v_y)^2}\right\|_1^2 + \left\|\sqrt{(\nabla_y v_x)^2 + (\nabla_y v_y)^2}\right\|_1^2 \quad (10)$$
$$+ \left\|\sqrt{(\nabla_{xy} v_x)^2 + (\nabla_{xy} v_y)^2}\right\|_1^2 + \left\|\sqrt{(\nabla_{yx} v_x)^2 + (\nabla_{yx} v_y)^2}\right\|_1^2$$

Regularization using the $\text{HVD}(\mathbf{v})$ imposes partial flow derivative continuity by minimizing their $\ell_1$ norms independently. Therefore, regularizing optical flow using Equation (10) preserves sharp horizontal, vertical, and diagonal edges. Moreover, including diagonal derivatives in a neighborhood around each pixel, in addition to the horizontal and vertical derivatives, imposes additional constraints on the unknown optical flow. Thus, the inclusion of these additional constraints in the $\text{HVD}(\mathbf{v})$ regularizer decreases staircase artefacts in the flat image regions. The $\ell_1$ minimization of each partial flow derivative separately favors a sparser solution than the solution acquired using TV. Thus, $\text{HVD}(\mathbf{v})$ is expected to require lesser measurements of the intensity derivatives than the TV regularizer for estimating unknown optical flow.

## V. OPTICAL FLOW ESTIMATION USING PROPOSED REGULARIZER

This section integrates the proposed regularizer and various optical flow data terms within a variational energy framework. The combined energy of the regularization and data terms is minimized using a modified version of a fast and convex algorithm [21]. The details of the proposed algorithm are also given.

Let $\mathbf{I_t}$, $\mathbf{I_x}$, and $\mathbf{I_y}$ be lexicographically vectorized temporal, horizontal, and vertical intensity derivatives, respectively. Additionally, $\mathbf{v_x}$ and $\mathbf{v_y}$ represent the vectorized horizontal and vertical optical flow components. The linearized optical flow constraint (OFC) can be mathematically represented as follows:

$$\left(\text{diag}(\mathbf{I_x}) \quad \text{diag}(\mathbf{I_y})\right)\begin{pmatrix}\mathbf{v_x}\\\mathbf{v_y}\end{pmatrix} = -(\mathbf{I_t}) \quad (11)$$

where $\text{diag}(\mathbf{I_x})$, $\text{diag}(\mathbf{I_y})$ represent diagonal matrices constructed using the horizontal and vertical intensity derivative vectors. For $A = [\text{diag}(\mathbf{I_x}) \; \text{diag}(\mathbf{I_y})]$, $\mathbf{v} = [\mathbf{v_x} \; \mathbf{v_y}]^T$, and $\mathbf{y_1} = -\mathbf{I_t}$, the matrix-vector form of the data term is given as

$$E_{data1}(\mathbf{v}) = \|A\mathbf{v} - \mathbf{y_1}\|_2^2 \quad (12)$$

The proposed regularizer given in Equation (9) can be written for the vectorized optical flow $\mathbf{v}$ as

$$E_{reg}(\mathbf{v}) = \text{HVD}(\mathbf{v}):$$
$$= \|\nabla_x \mathbf{v}\|_1^2 + \|\nabla_y \mathbf{v}\|_1^2 + \|\nabla_{xy}\mathbf{v}\|_1^2 + \|\nabla_{yx}\mathbf{v}\|_1^2 \quad (13)$$

The joint variational energy $E(\mathbf{v})$ of Equations (12) and (13) can be expressed as

$$E(\mathbf{v}) = E_{data1}(\mathbf{v}) + E_{reg}(\mathbf{v}) = \|A\mathbf{v} - \mathbf{y_1}\|_2^2$$
$$+ \lambda \|\nabla_x \mathbf{v}\|_1^2 + \|\nabla_y \mathbf{v}\|_1^2 + \|\nabla_{xy}\mathbf{v}\|_1^2 + \|\nabla_{yx}\mathbf{v}\|_1^2. \quad (14)$$

When the brightness constancy assumption is violated, the OFC based data term cannot accurately match pixel intensities across video frames. In this case, intensity gradient constancy performs better so we can develop a data matching term based on the constancy of the intensity gradients. The intensity gradient constancy assumption (GCA) can be expressed as

$$\nabla I(x, y, t) - \nabla I(x - v_x dt, y - v_y dt, t - dt) = 0. \quad (15)$$

We linearize the GCA to develop a data term that can handle intensity change. The linearization of Equation (15) results

$$\begin{pmatrix} I_{xx} v_x + I_{xy} v_y \\ I_{yx} v_x + I_{yy} v_y \end{pmatrix} = -\begin{pmatrix} I_{xt} \\ I_{yt} \end{pmatrix} \quad (16)$$

and

$$\begin{pmatrix} I_{xx} & I_{xy} \\ I_{yx} & I_{yy} \end{pmatrix}\begin{pmatrix} v_x \\ v_y \end{pmatrix} = -\begin{pmatrix} I_{xt} \\ I_{yt} \end{pmatrix}. \quad (17)$$

When $\mathbf{I_{xx}}$, $\mathbf{I_{xy}}$, $\mathbf{I_{yx}}$, and $\mathbf{I_{yy}}$ represent second-order intensity derivatives in a vectorized form and $\mathbf{I_{xt}}$, $\mathbf{I_{yt}}$ denote spatiotemporal intensity derivatives, the linearized intensity GCA can be given as

$$\begin{pmatrix} \text{diag}(\mathbf{I_{xx}}) & \text{diag}(\mathbf{I_{xy}}) \\ \text{diag}(\mathbf{I_{yx}}) & \text{diag}(\mathbf{I_{yy}}) \end{pmatrix}\begin{pmatrix} \mathbf{v_x} \\ \mathbf{v_y} \end{pmatrix} = -\begin{pmatrix} \mathbf{I_{xt}} \\ \mathbf{I_{yt}} \end{pmatrix} \quad (18)$$

Diagonal matrices: $\text{diag}(\mathbf{I_{xx}})$, $\text{diag}(\mathbf{I_{xy}})$, $\text{diag}(\mathbf{I_{yx}})$ and $\text{diag}(\mathbf{I_{yy}})$, in Equation (18) are constructed using the vectors $(\mathbf{I_{xx}})$, $(\mathbf{I_{xy}})$, $(\mathbf{I_{yx}})$ and $(\mathbf{I_{yy}})$, respectively. For:

$$B = \begin{pmatrix} \text{diag}(\mathbf{I_{xx}}) & \text{diag}(\mathbf{I_{xy}}) \\ \text{diag}(\mathbf{I_{yx}}) & \text{diag}(\mathbf{I_{yy}}) \end{pmatrix} \quad \text{and} \quad \mathbf{y}_2 = -\begin{pmatrix} \mathbf{I_{xt}} \\ \mathbf{I_{yt}} \end{pmatrix};$$

the intensity gradient constancy assumption-based data term is given as

$$E_{data2}(\mathbf{v}) = \|B\mathbf{v} - \mathbf{y}_2\|_2^2 \qquad (19)$$

Note that both Equations (12) and (19) have similar form. The $E_{\text{data2}}(\mathbf{v})$ is computationally more intense than $E_{\text{data1}}(\mathbf{v})$ because it consists of two equations.

When the brightness variation is slow across video frames, we can model the data term as the following function of the image intensity as

$$I(x, y, t) - I(x - v_x dt, y - v_y dt, t - dt)$$
$$= d(x, y, t)I(x, y, t) + c(x, y, t). \qquad (20)$$

where the offset $c(x, y, t)$ permits the brightness change and the multiplier $d(x, y, t)$ allows variation in the image contrast. The aforementioned equation is an example of a generalized dynamic image model (GDIM) whose linearized version is expressed as

$$I_x v_x + I_y v_y + I_t = dI + c. \qquad (21)$$

Let the vectorized $I$, $d$ and $c$ be $\mathbf{I}$, $\mathbf{d}$ and $\mathbf{c}$, respectively. The vectorized form of Equation (21) is given as

$$\begin{pmatrix} \text{diag}(\mathbf{I_x}) & \text{diag}(\mathbf{I_y}) & -\text{diag}(\mathbf{I}) & -I_n \end{pmatrix} \begin{pmatrix} \mathbf{v_x} \\ \mathbf{v_y} \\ \mathbf{d} \\ \mathbf{c} \end{pmatrix} = -\mathbf{I_t} \qquad (22)$$

where $I_n$ represents an $n \times n$ identity matrix. Let $C = \begin{pmatrix} \text{diag}(\mathbf{I_x}) & \text{diag}(\mathbf{I_y}) & -\text{diag}(\mathbf{I}) & -I_n \end{pmatrix}$ and $\mathbf{u} = (\mathbf{v}_x \ \mathbf{v}_y \ \mathbf{d} \ \mathbf{c})^T$. The GDIM based data term can now be given as

$$E_{data3}(\mathbf{v}, \mathbf{d}, \mathbf{c}) = \|C\mathbf{u} - \mathbf{y}_1\|_2^2 \qquad (23)$$

The data term $E_{\text{data3}}(\mathbf{v}, \mathbf{d}, \mathbf{c})$ has parameters that capture the change in the image brightness. These parameters have to be computed along with $\mathbf{v}$.

### A. ALGORITHMIC DETAILS

The data terms in Equations (12), (19), and (23) have the same form. Thus, these data terms can be incorporated in the energy minimization framework and the same optimization algorithm can be employed to minimize them. We employ the NESTA method, which is an effective approach [21], to estimate the optical flow based on Equation (14). NESTA approximates the $\ell_1$ norm with a differentiable Huber norm; thus, it can handle non-smooth functions. The Huber norm of $x$ is given as

$$\|x\|_\epsilon = \begin{cases} \dfrac{x^2}{2\epsilon}, & \text{if } |x| \leq \epsilon \\ |x| - \dfrac{\epsilon}{2}, & \text{otherwise} \end{cases} \qquad (24)$$

Its derivative is given as

$$\frac{\partial}{\partial x}\|x\|_\epsilon = \frac{x}{\max(|x|, \epsilon)}. \qquad (25)$$

We replace the $\ell_1$ norm by the differentiable Huber norm in Equation (14), which results in the joint data and the regularization energy $E(\mathbf{v})$:

$$E(\mathbf{v}) = \min_{\mathbf{v}} \|A\mathbf{v} - \mathbf{y}\|_\epsilon + \lambda \left\{ \begin{array}{l} \left\|\sqrt{(\nabla_x v_x)^2 + (\nabla_x v_x)^2}\right\|_\epsilon + \\ \left\|\sqrt{(\nabla_y v_x)^2 + (\nabla_y v_x)^2}\right\|_\epsilon + \\ \left\|\sqrt{(\nabla_{xy} v_x)^2 + (\nabla_{xy} v_x)^2}\right\|_\epsilon + \\ \left\|\sqrt{(\nabla_{yx} v_x)^2 + (\nabla_{yx} v_x)^2}\right\|_\epsilon \end{array} \right\}. \qquad (26)$$

We compute the minimum of Equation (26) at iteration $k$ by using an iterative scheme as

$$\frac{\partial E(\mathbf{v}^k)}{\partial \mathbf{v}^k} = \partial_\mathbf{v} E(\mathbf{v}^k) = \frac{A^T(Av^k + y)}{\max(\epsilon, |Av^k + y|)}$$

$$+\lambda \left\{ \begin{array}{l} \nabla_x^T \dfrac{\nabla_x v^k}{\max(\epsilon, |\nabla_x v^k|)} + \\ \nabla_y^T \dfrac{\nabla_y v^k}{\max(\epsilon, |\nabla_x v^k|)} + \\ \nabla_{xy}^T \dfrac{\nabla_{xy} v^k}{\max(\epsilon, |\nabla_{xy} v^k|)} + \\ \nabla_{yx}^T \dfrac{\nabla_{yx} v^k}{\max(\epsilon, |\nabla_{yx} v^k|)} \end{array} \right\}. \qquad (27)$$

At each iteration, the algorithm computes auxiliary variables $\mathbf{p}^k$ and $\mathbf{q}^k$ from $\partial_{\mathbf{v}^k} E(\mathbf{v}^k)$. The algorithm uses Lipschitz continuity; thus, a Lipschitz constant $L = 16\lambda/\epsilon$ is used in Equation (27) to compute auxiliary variables.

To handle large displacements of pixels, we apply the proposed algorithm on a Laplacian image pyramid as in [4, 9]. The Laplacian pyramid provides a multiresolution representation of an image at its different levels. It has successive video frames at various resolutions to enable a coarse to fine estimation of optical flow. We initialize the optical flow to 0 at the lowest resolution level. The computed optical flow vector $\mathbf{v}$ at each pyramid level is then carried over to the subsequent finer level as an initial guess $\mathbf{v}^0$. The algorithm operates for a set number of iterations at each pyramid level or until it converges. Table 1 provides a summary of the algorithm.

Table 1: The proposed coarse to fine algorithm for estimating optical flow vectors from successive image sequences.

**Initialization**: $\mathbf{v}^0 = \begin{bmatrix} \mathbf{v}_x^0 & \mathbf{v}_y^0 \end{bmatrix}^T = 0$.
**for** level=, ... , 1
  **if**
    $\mathbf{v} = \mathbf{v}^0$,
  **else**
    $\mathbf{v}^0 =$ up-sample$(\mathbf{v})$,
  **end**
  Set iteration index $k = 1$,
  **While** ($k \leq iter_{max}$ & without convergence)
    1. calculate $\partial_{\mathbf{v}^k} E(\mathbf{v}^k)$ using (27),
    2. determine $\gamma^k = \frac{1}{2}(k+1)$ and $\tau^k = \frac{2}{k+3}$,
    3. compute $\mathbf{p}^k = \mathbf{v}^k - \frac{1}{L} \partial_{\mathbf{v}^k} E(\mathbf{v}^k)$,
    4. compute $\mathbf{q}^k = \mathbf{v}^0 - \frac{1}{L} \sum_i^k \gamma^i \partial_{\mathbf{v}^i} E(\mathbf{v}^i)$,
    5. update $\mathbf{v}^k = \tau^k \mathbf{p}^k + (1 - \tau^k) \mathbf{q}^k$,
  $k = k + 1$,
  **end while**
  $\mathbf{v} = \mathbf{v} + \mathbf{v}^0$.
**end for**

## VI. RESULTS AND DISCUSSION

In this section, we assess the capabilities of the proposed method to estimate optical flow and analyze the experimental outcomes. We describe the setup for experiments and the details of the dataset used to evaluate the proposed method in subsection A. The proposed regularizer incorporated with different data terms is evaluated in subsection B. We also conducted experiments to quantity the proposed method's effectiveness in estimating optical flow using a reduced number of measurements in subsection C. We compared the performance of the proposed method against several existing approaches in subsection D.

### A. EXPERIMENTAL SETUP AND DATASET

We utilized a Laplacian pyramid that down-samples the video sequences by a factor of 0.70 to manage large displacements. We carefully modified the parameter λ to achieve accurate results during our experiments, keeping it within the range of [1×10⁻³  1×10⁻¹]. The parameter $\epsilon$ in Huber norm significantly influences the algorithm's convergence. In our experiments, we set $\epsilon = 0.01$, leading to reasonable accuracy and convergence. On average, convergence of the algorithm occurred after a few hundred iterations, although we set a maximum of 500 iterations for the algorithm. We have used a color scheme to indicate the amplitude and direction of the estimated optical flow. This color scheme is given in Figure 2. We expressed quantitative findings using the mean-end-point-error (MEPE) metric, which is calculated using the ground truth optical flow as:

$$\text{MEPE} = \frac{1}{n} \sum_{i=1}^{n} \sqrt{\left[ (v_x)_i - (v_{xGT})_i \right]^2 + \left[ (v_y)_i - (v_{yGT})_i \right]^2} \quad (28)$$

where $\mathbf{v}_{GT} = \begin{bmatrix} \mathbf{v}_{xGT} & \mathbf{v}_{yGT} \end{bmatrix}^T$ represents the ground truth flow, and $n$ denotes the number of pixels in a single video frame.

We have used the Middlebury dataset [22], which is designed for the evaluation and benchmarking of optical flow algorithms. For evaluation purposes, the dataset provides ground-truth optical flow for eight training video sequences. These sequences fall into different categories. Some sequences such as *RubberWhale, Hydrangea,* and *Dimetrodon* have hidden texture. *Grove2*, *Grove3*, *Urban2*, and *Urban3* are synthetic, meaning they are generated artificially rather than being captured from real-world scenarios. The dataset includes stereo sequences such as *Venus*. Stereo sequences are essential for evaluating how well optical flow algorithms handle depth information.

### B. VARIOUS DATA MATCHING TERMS WITH THE PROPOSED REGULARIZER

We utilized the synthetically generated training video sequences from the Middlebury dataset for these experiments. These sequences' ground truth optical flow is also available. The video sequences involve either camera motion or motion of rigid objects. Figure 2 displays the optical flow estimation results when we use the OFC based data term with the proposed regularizer. The estimated flows are shown as color plots. Notably, Figure 2 (b) and (c) demonstrate minimal visible disparity between the estimated optical flows and ground truth, showcasing the preservation of motion boundaries by the proposed method. Figure 2 (d) highlights errors occurring at occluded boundaries, particularly noticeable in the Urban2 and Venus sequences due to occlusions at image boundaries.

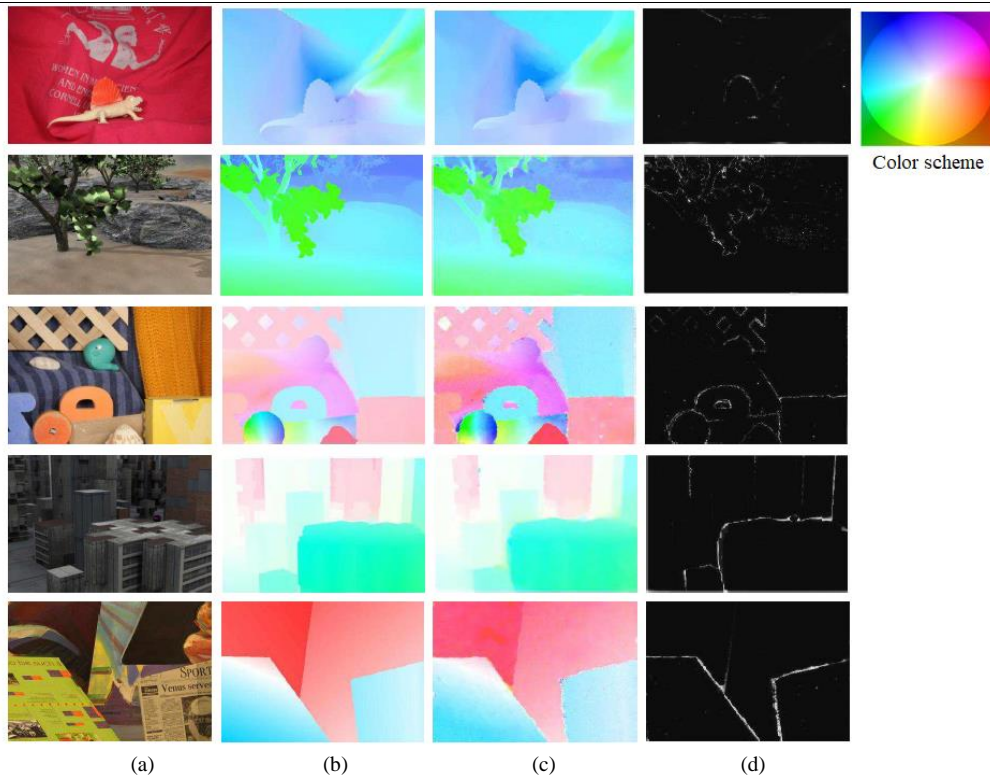

(a) (b) (c) (d)

Figure 2: Optical flow estimation with the suggested OFC-based approach. Every row represents a distinct image sequence. (a) Image frame, (b) ground truth flow, (c) estimated optical flow, and (d) error among the estimated optical flow and ground truth. The color scheme shows the flow vectors' amplitude and direction.

Table 2 presents quantitative MEPE results for the mentioned video sequences. We applied five distinct data terms in conjunction with the proposed regularizer. Data terms utilizing the OFC and OFC-based neighborhood-dependent constraint (NDC) performed well when the assumption of brightness constancy was not violated. On the other hand, three other data terms, namely the GCA, GDIM, and the OFC for the normalized data (N-OFC), effectively handled brightness changes. For sequences like Rubberwhale and Urban3, which experienced brightness changes, N-OFC, GCA, and GDIM exhibited superior performance in terms of MEPE compared to the other data terms.

We also used real-world sequences to analyze the capabilities of the proposed method. Through experimentation, we found λ=0.005 to be the optimal value for the used sequences. To enhance robustness against brightness changes, we pre-processed real video sequences. Specifically, we applied a Gaussian filter of size 9×9 and a σ=1 to smooth these sequences. The resulting smoothed sequences were subtracted from the raw sequences, and the resultant sequences were then employed for optical flow estimation. All other algorithm settings remained consistent.

Table 2: The proposed HVD regularizer with different data terms, MEPE results on the Middlebury dataset.

|  | Data Terms | | | | |
| --- | --- | --- | --- | --- | --- |
| **Sequence** | NDC [4] | N-OFC [30] | Proposed regularizer with OFC | Proposed regularizer with GCA | Proposed regularizer with GDIM |
| Dimetrodon | 0.16 | **0.12** | 0.15 | 0.17 | 0.21 |
| Grove2 | 0.21 | 0.15 | **0.12** | 0.14 | 0.18 |
| Grove2 | 0.52 | 0.48 | **0.41** | 0.55 | 0.43 |
| Hydrangea | 0.22 | 0.25 | **0.16** | 0.18 | 0.21 |
| Rubberwhale | 0.18 | 0.15 | 0.16 | **0.10** | **0.10** |
| Urban2 | 0.54 | 0.58 | 0.41 | 0.48 | **0.43** |
| Urban3 | 0.62 | 0.49 | 0.65 | **0.45** | 0.51 |
| Venus | 0.34 | 0.29 | 0.28 | **0.26** | 0.32 |

Figure 3 demonstrates the estimation of the optical flow for real-world sequences: Figure 3 (a)-(d) *Walking*, *Dumptruck*, *Flower garden*, and *Camera motion*. In the *Flower garden* image sequence, a moving camera captures a scene. *Walking* and *Dumptruck* are taken from the Middlebury test dataset. In *Dumptruck*, the camera is stationary, whereas both the camera and a person are in motion in the *Walking* sequence. Although ground truth data is not available for the mentioned real sequences. In *Camera motion* sequence, a moving camera records a complex multi-object scene. The results are convincing, depicting the preservation of motion boundaries in all four sequences.

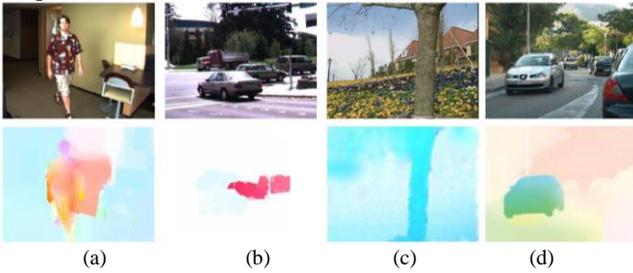

Figure 3: Optical flow estimation on real-world image sequences. The top row displays the image frames, while the estimated optical flows are presented in the bottom row.

## C. EFFECT OF REDUCED MEASUREMENTS ON OPTICAL FLOW ESTIMATION

These experiments assess the capabilities of the method proposed in Equation (14) under a small number of intensity derivative measurements. Three distinct schemes were employed for intensity derivative selection:

- Random selection of derivatives
- Significant derivative magnitude selection
- Combined significant and random derivatives selection

For $m$ selected pixel in an $n$-pixel image frame, let us define the measurement ratio as $m/n$. We evaluate the performance for various values of the measurement ratio. Notably, when $m/n=1$, full measurements were utilized, implying no reduction. For each type of selection, we gradually increased the measurements, and plotted the MEPE alongside each measurement ratio. For random and combined selection schemes, the algorithm was run 5 times for each measurement ratio, and the average of the results was considered in the end. We set the parameter $\lambda$ to 0.01.

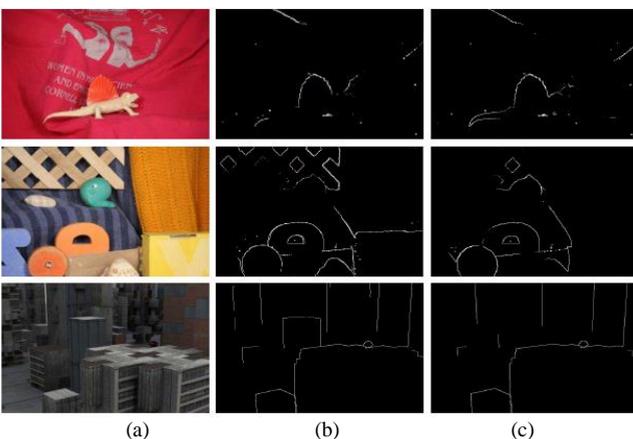

Figure 4: Depiction of the ground truth flow sparsity levels. (a) Image frame, the gradient magnitude of (b) $\mathbf{v}_{x\text{GT}}$ and (c) $\mathbf{v}_{y\text{GT}}$ ground truth flow vectors.

We also conducted a comparison between the percentage of intensity derivative measurements required to correctly estimate optical flow and the sparsity of the partial flow derivatives. To do this, we computed magnitude maps using partial flow derivatives and the ground truth flow gradient. For the purpose of binarizing these maps, Otsu's global thresholding was applied. Subsequently, we counted the number of non-zero components in each result to compute the sparsity of these maps. We utilized Middlebury's image sequences for these experiments due to the availability of the ground truth data. Figure 4 shows the magnitude maps representing the gradient of the ground truth flows for three different video sequences. The maps represent motion boundaries $\mathbf{v}_{x\text{GT}}$ and $\mathbf{v}_{y\text{GT}}$, and are depicted in Figure 4 (b) and (c), respectively. It is evident that a small percentage of the total number of pixels form motion boundaries.

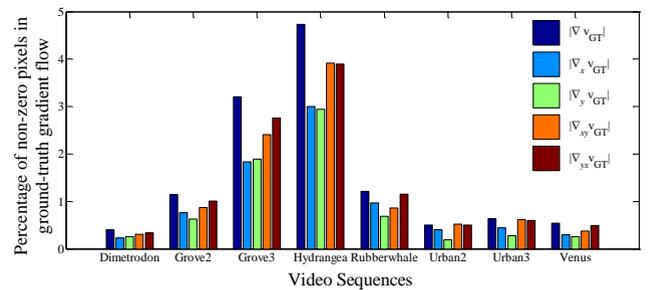

Figure 5: The nonzero pixels in percentage computed from the derivative maps of the Middlebury ground truth flow vectors.

Figure 5 demonstrates that about 5% or less number of pixels constitute nonzero partial derivative magnitudes and gradient magnitude of optical flow in these image frames. It's important to note that the partial flow derivatives' magnitude maps: $|\boldsymbol{\nabla}_x \mathbf{v}_{\text{GT}}|$, $|\boldsymbol{\nabla}_y \mathbf{v}_{\text{GT}}|$, $|\boldsymbol{\nabla}_{xy} \mathbf{v}_{\text{GT}}|$, and $|\boldsymbol{\nabla}_{yx} \mathbf{v}_{\text{GT}}|$, exhibit greater sparsity compared to the magnitude map of the gradient flow $|\nabla \mathbf{v}_{\text{GT}}|$.

Figure 6 shows the MEPE against the measurement ratio for 3 different types of intensity derivative selections. We see an increase in error for the majority of the video sequences when the significant derivative selection is used for a measurement ratio of 0.4 or lower. Relying only on intensity derivatives with significant magnitude may fail to detect moving objects that have weak edges when a scene has moving objects that have strong edges alongside others with weak edges. Thus, this leads to a higher MEPE values.

The error does not increase noticeably when we utilize a random selection of intensity derivatives even for $m/n=0.1$, except for the Rubberwhale and Grove3 sequences. In the Grove3 sequence, which involves small objects in motion, a 0.1 measurement ratio of random selection of derivatives misses detail of small objects, causing an increase in error (see Figure 6 (b). On the other hand, in the Rubberwhale sequence that contains motion of different objects, the error increases for $m/n=$ 0.1 as in Figure 6 (d).

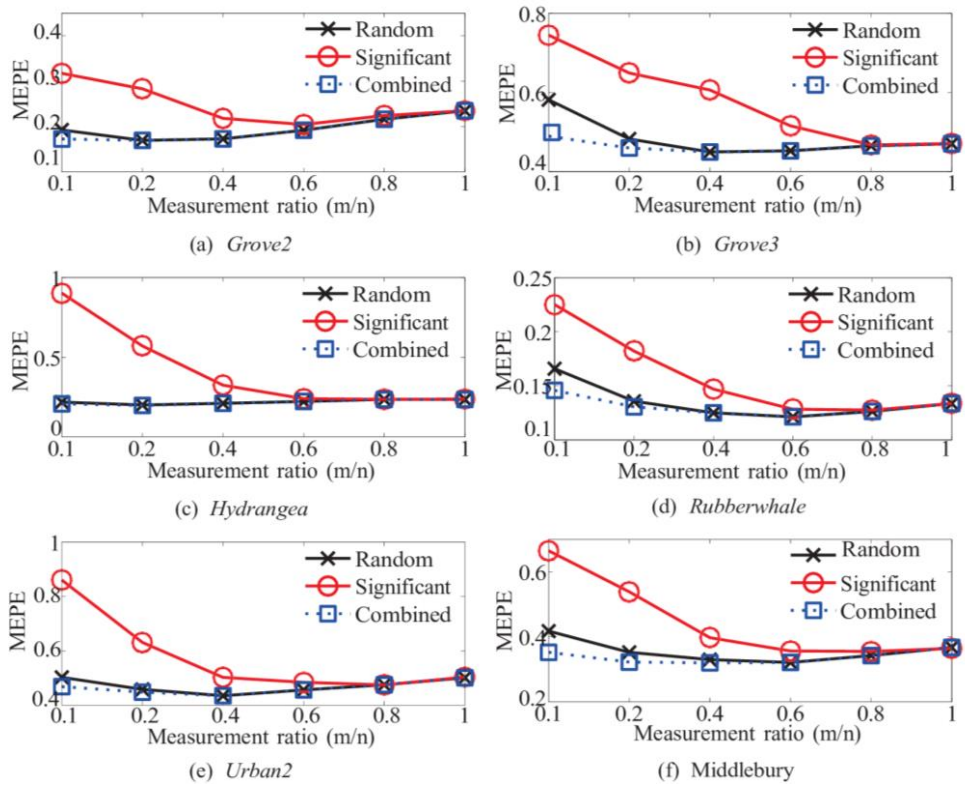

Figure 6: The MEPE against the measurement ratio for 3 different selection schemes across various sequences. The Middlebury dataset's MEPE averaged over eight videos is given in (f).

The *Hydrangea* sequence has 2 distinctive motions of the bush in the foreground and the background. Thus, optical flow is correctly estimated for 0.1 measurement ratio. In contrast to random only or significant only sensing, the combined sensing scheme computes the optical flow correctly for the measurement ratio as small as $m/n$=0.1.

### D. COMPARISON WITH OTHER METHODS

In these experiments, we evaluate the proposed HVD regularizer-based method in comparison to the various optical flow methods. The methods included in this comparison are the anisotropic TV method TV-L1-improved [2], the image adaptive method Adaptive [29], the adaptive TV and a neighborhood data term-based method Ad-TV-NDC [4], the nonlocal TV method Classic + NL [7], and the optical flow refinement method RF grid + mix [31]. The refinement method uses the local flow variance and temporal evolution of the Kanade–Lucas–Tomasi (KLT) residuals to control the refinement by reliability measures and color local homogeneity.

Furthermore, to evaluate the impact of adaptive regularization, image adaptive regularization was incorporated into the proposed method. We made use of the adaptive weighting method that was proposed in [4], and we refer to this method as Proposed Adaptive HVD. We carry out the comparison for reduced set of measurements, focusing on the combined sensing scheme, which has shown improved performance. In this combined sensing approach, we utilized $0.05n$ significant intensity derivatives, and the remaining measurements were selected randomly.

Table 3 compares the proposed method with existing methods computed for 8 image sequences of the Middlebury dataset. It is evident that Adaptive, Ad-TV-NDC, and TV-L1-improved exhibit higher MEPE compared to the other three methods. In the majority of videos, Proposed Adaptive HVD demonstrates superior performance over other methods.

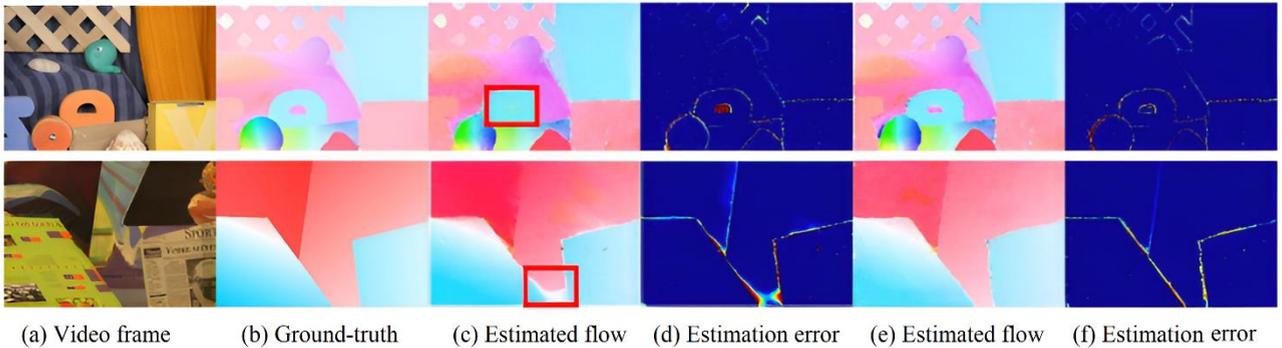

(a) Video frame  (b) Ground-truth  (c) Estimated flow  (d) Estimation error  (e) Estimated flow  (f) Estimation error

Figure 7: The computed optical flow and the resulting error for a measurement ratio $m/n = 0.2$, employing a combined sensing of intensity derivatives. (a) Image frame, (b) the ground truth flow, (c)-(d) Adaptive [29] method's estimated flow and resulting error, (e)-(f) Proposed HVD method's estimated flow and error.

Table 3: The MEPE results of various methods on Middlebury dataset, including the proposed method.

| Methods →<br>Sequences ↓ | TV-L1 Improved [2] | Adaptive [29] | Classic + NL [7] | Ad-TV-NDC [4] | RF grid + mix [31] | Proposed HVD | Proposed Adaptive HVD |
|---|---|---|---|---|---|---|---|
| Dimetrodon | 0.18 | 0.16 | **0.13** | 0.17 | 0.15 | 0.15 | 0.14 |
| Grove2 | 0.19 | 0.16 | 0.16 | 0.17 | 0.15 | 0.15 | **0.13** |
| Grove3 | 0.62 | 0.51 | 0.48 | 0.58 | 0.55 | 0.41 | **0.39** |
| Hydrangea | 0.25 | 0.21 | 0.15 | 0.25 | 0.20 | **0.16** | 0.17 |
| Rubberwhale | 0.21 | 0.18 | 0.15 | 0.16 | **0.11** | 0.12 | **0.11** |
| Urban2 | 0.57 | 0.49 | 0.39 | 0.46 | 0.38 | 0.41 | **0.37** |
| Urban3 | 0.77 | 0.64 | 0.52 | 0.68 | **0.49** | 0.56 | 0.59 |
| Venus | 0.38 | 0.33 | 0.29 | 0.32 | 0.29 | 0.28 | **0.26** |
| **Average** | 0.39 | 0.34 | 0.28 | 0.33 | 0.29 | 0.25 | **0.24** |

Figure 7 shows the estimated flows and errors for the Proposed HVD and Adaptive methods using $m/n= 0.2$. In Figure 7 (c), the optical flow estimated by the Adaptive method fails to estimate motion of small objects in the highlighted part of the *Rubberwhale* video sequence in the first row. An excessive blurring in the lower to middle section of the *Venus* image sequence in the second row can also be observed. Conversely, the proposed method avoids these issues, as in Figure 7 (e).

Figure 8 illustrates the MEPE results on different measurement ratios for various methods on the Middlebury training dataset. A combined sensing scheme is used for different measurement ratios. The error is almost constant for Adaptive-HVD and Proposed HVD, even for $m/n$=0.1. Notably, incorporating adaptive regularization alongside the HVD regularizer does not significantly enhance error performance, suggesting that the anisotropic HVD regularizer effectively preserves sharp motion boundaries without adaptive weighting. Classic + NL generally yields good performance since it uses multiple nonlocal pixels for the construction of the data term.

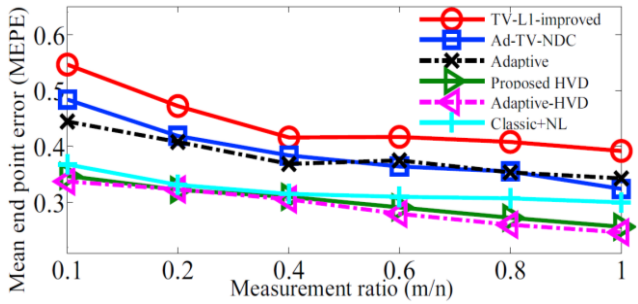

Figure 8: The average MEPE against the measurement ratio, calculated over all videos in the Middlebury training dataset. The comparison involves the following methods: TV-L1-improved [2], Adaptive [29], Ad-TV-NDC [4], Classic + NL [7], Proposed using combined sensing (Proposed HVD), and the proposed adaptive using combined sensing (Adaptive-HVD).

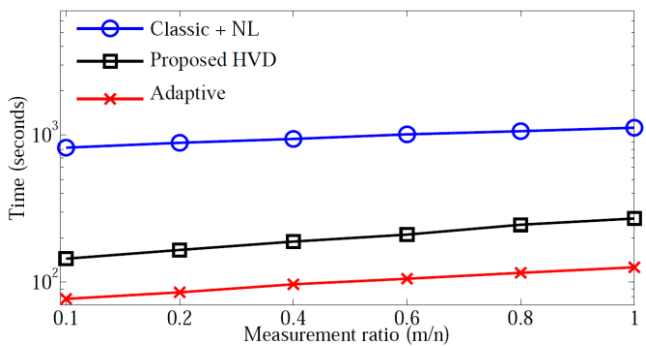

Figure 9: Comparison of the run time against different sets of measurement ratios for Proposed HVD, Adaptive, and Classic + NL methods.

We also compared the computation time of the Adaptive, Classic + NL, and Proposed HVD methods for different number of measurements in Figure 9. The experiments demonstrate that using 20% intensity derivative measurements in Adaptive reduces the computational time by up to 30%. Additionally, the proposed method achieves approximately a 20% time gain for the similar quantity of measurements. While Classic + NL performs comparable to the proposed method, it is significantly slower. Classic + NL incurs the most time due to the higher computational cost associated with the minimization of nonlocal terms in this method.

## VII. CONCLUSION

This paper introduces a novel regularizer for dense optical flow estimation, leveraging the sparsity of the derivatives of motion vectors in the spatial domain. The regularizer enforces partial flow derivative continuity within small neighborhoods around each flow vector and minimizes them separately. Three distinct data terms were employed in conjunction with the proposed regularizer. When OFC was utilized as a data term, it was found inadequate for handling variations in brightness within digital videos. To address optical flow estimation under changing brightness conditions, two data terms were incorporated: GCA and GDIM. Experimental evaluations were conducted on synthetic and real video sequences to estimate optical flow. Results demonstrated that the proposed method did not cause blurring across sharp motion boundaries, whether they are horizontal, vertical, or diagonal. Despite its computational efficiency, the proposed method's performance is on par with nonlocal total variation. We employed three different spatiotemporal intensity derivative sensing schemes: significant, random, and combined sensing for optical flow estimation from a small number of intensity derivative measurements. Experimental results indicated that the combined sensing of intensity derivatives outperformed both random-only and significant-only sensing. By employing a combined intensity derivative sensing scheme with the proposed regularizer, it was demonstrated that optical flow estimation with only 10% of total measurements could be achieved without significantly compromising accuracy.